\documentclass{llncs}
\usepackage{llncsdoc}

\usepackage{caption}
\usepackage{float}
\usepackage[caption = false]{subfig}
\usepackage{graphicx}
\usepackage[spanish]{babel}
\decimalpoint
\usepackage[utf8]{inputenc}
\usepackage{array}
\usepackage{algorithm}
\usepackage{algpseudocode}
\usepackage{amsmath}
\usepackage{url}

\begin{document}
\renewcommand{\tablename}{Table}
\renewcommand{\figurename}{Fig.}
\floatname{algorithm}{Algorithm}
\renewcommand{\algorithmicrequire}{\textbf{Input:}}
\renewcommand{\algorithmicensure}{\textbf{Output:}}

\title{Evolutionary optimization of contexts for phonetic correction in speech recognition systems}
\titlerunning{Evolutionary optimization of contexts for correction in ASR systems}

\author{Rafael {Viana Cámara}, Diego {Campos Sobrino}, Mario {Campos Soberanis}}
\authorrunning{R. Viana et al.}

\institute{SoldAI Research, Calle 22 No. 202-O, García Ginerés, 97070 Mérida, México \\
\email{\{rviana,dcampos,mcampos\}@soldai.com} \\
}

\maketitle

\begin{abstract}
Automatic Speech Recognition (ASR) is an area of growing academic and commercial interest due to the high demand for applications that use it to provide a natural communication method. It is common for general purpose ASR systems to fail in applications that use a domain-specific language. Various strategies have been used to reduce the error, such as providing a context that modifies the language model and post-processing correction methods. This article explores the use of an evolutionary process to generate an optimized context for a specific application domain, as well as different correction techniques based on phonetic distance metrics. The results show the viability of a genetic algorithm as a tool for context optimization, which, added to a post-processing correction based on phonetic representations, can reduce the errors on the recognized speech.\\

\end{abstract}

\section{Introduction}
Automatic speech recognition (ASR) systems are of great relevance in academic and business environments due to the ease of interaction they offer. There has been a growing interest in investigating these systems, which have migrated from probabilistic models to deep neural network systems \cite{Becerra2016} that have become the standard for professional audio-to-text transformation applications. Deep neural network systems for audio-to-text transformation often use an acoustic model to perform recognition at a first level and are later passed to language models for correction \cite{Droppo2010}. Commercial services generally operate as a black box, making it difficult for the user to modify language models.

While ASRs generally perform well, they often run into problems when used to recognize specific language domains, so post-processing techniques become relevant \cite{Bassil2012}.

Many of the post-processing and correction tasks of these systems use a context, understood as a set of words, phrases, and expressions related to the particular domain that it is desired to recognize. Some of them have mechanisms to provide a context with which they improve the recognition of certain words and phrases. However, in many cases, it is not enough to significantly improve their performance.

Two particularly interesting topics are the generation of contexts and the phonetic representation for correction. Research has been carried out in this regard; however, there is a lack of experimentation related to the optimal context generation and phonetic representation's joint operation.

This article presents a method for generating contexts using genetic algorithms to correct the output of the \emph{Google} speech-to-text processing system. Next, the error correction process's comparison of different critical strategies is carried out: representation of the sentence to be corrected, candidate selection, and comparison metrics.

The article is structured as follows: Section 2 describes the background to the problem and related work; Section 3 presents the methodology used for the investigation; In section 4, the experimental work carried out is described, the results of which are shown in section 5 and finally in section 6 the conclusions are provided along with some ideas to develop as future work.

\section{Background}
The error correction algorithms in ASR systems have been approached from different perspectives, including phonetics. Kondrak \cite{kondrak2003} proposes an algorithm to calculate a metric of phonetic similarity between segments using multivalued articulatory phonetic characteristics. The Kondrak algorithm combines sets of edit operations and local and semi-global alignment models to calculate a set of near-optimal alignments.

Pucher et al. \cite{Pucher2007} present word confusion matrices using different measures of phonetic distance. The metrics presented are based on the minimum editing distance between phonetic transcriptions and the distances between hidden Markov models. His research shows a correlation between edit distance and word confusion in ASR systems, so these types of corrections become useful for rectifying recognition errors.

In \cite{Bae2012} the problem of using the editing distance to compare strings in languages like Korean, where characters represent syllables instead of letters, is highlighted. This is reflected in the fact that substituting one syllable for another provides the same value regardless of the difference between its letters. The traditional solution uses hybrid metrics between characters and syllables; however, the authors argue that this approach does not satisfactorily solve the problem, so they propose an editing distance based on phonemes as a solution.

Droppo and Acero \cite{Droppo2010} use the phonetic editing distance to incorporate a third element of correction to ASR systems. They incorporate this distance to learn the relative probability of phonetic recognition strings, given an expected pronunciation. This strategy considers the context of the transcripts, changing the probability of correction depending on the words before and after.

Bassil and Semaan \cite{Bassil2012} employ a post-processing strategy for error correction in ASR systems. The presented method for detecting word errors uses a candidate generation algorithm and a context-sensitive error correction algorithm. The authors report a significant reduction in system errors.

In \cite{Campos2018} phonetic correction strategies are used to correct the errors generated by an ASR system. In the cited work, the system's transcript is transformed into a representation in International Phonetic Alphabet (IPA) format. A sliding window algorithm is used to select candidate phrases for correction according to the words provided, in context and distance to its phonetic representation in IPA format. The authors report an improvement in 30 \% of the phrases recognized by the \emph{Google} service.

An important component for the phonetic correction algorithm is the context used to construct the candidate phrases, so solutions capable of finding optimal configurations among vast search spaces are needed.

Genetic algorithms are stochastic search algorithms based on biological evolution principles and emulate the process through genetic operators applying recombination, mutation, and natural selection in a population \cite{ga2018,knapsack2019}. They have been applied to solve complex combinatorial problems, and the results show that they constitute a powerful and efficient strategy when used correctly. \cite{knapsack2019}.

These types of algorithms have been used to analyze a large number of problems, including knapsack \cite{knapsack2019}, process scheduling problems, the traveling salesperson \cite{ga2018}, search for functions for symbolic regression \cite{symbolicga2019}, Gaussian kernel functions for sentiment analysis \cite{gaussian2019}, among others.

When using genetic algorithms to solve a problem, possible solutions are expressed as a chain of symbols called a chromosome, where each symbol is a gene. From an initial generation of individuals, the processes of selection, mutation, recombination, and evaluation are iteratively executed, combining the individuals' genes to produce new variations. Each individual is evaluated according to a function called \emph{fitness} that describes how well they perform to solve the problem.

\section{Methodology}
The correction algorithm uses the components of a context (set of words and phrases belonging to the application domain) to detect possible errors in the recognition and correct the transcript of an automatic speech recognition system. It comprises three main elements: phonetic representation, candidate phrase generator for correction, and editing distance metric. As a measure of evaluation of the results, the WER metric (\textit{Word Error Rate}) was used, which is defined as follows:

\begin{equation}
WER = \dfrac {S + D + I} {N}
\end{equation}

where $S$ is the number of substitutions, $D$ the number of deletions, $I$ the number of insertions required to transform the hypothetical phrase into the actual phrase, and $N$ the number of words in the actual phrase.

\subsection{Phonetic transcription}
Phonetic transcription is a system of graphic symbols that represent the sounds of human speech. It is used as a convention to avoid the peculiarities of each written language and to represent those languages without a written tradition \cite{Hualde2005}. We use as phonetic representations: plain text, IPA, Double Metaphone (DM), and a variant of Double Metaphone with vowels (DMV).

The IPA is a phonetic notation system based on the Latin alphabet, used as a standardized representation of the sounds of the spoken language \cite{MacMahon,Wall}. Metaphone is a phonetic algorithm that is in charge of indexing words by their pronunciation from the English language \cite{Lawrence1990}. The DM algorithm is an improved version of the Metaphone algorithm, which returns a representation of the letters' sound in the string when the text is spoken and omits the vowels. The DM has often been used to represent the English language; however, vowel sounds are of importance in Spanish because they serve the Spanish speaker to link words that end in consonant groups \cite{ChelaFlores2006}, so we developed a variant of the DM which adds the vowels that are removed in the original algorithm.

\subsection{Candidate Sentence Generation Algorithms}
During the phonetic correction process, the search for candidate phrases generates segments of the input string that will be contrasted employing a distance metric with the words and phrases in the context. A candidate phrase is one that is similar to one of the phrases in the context and that may contain an error in the ASR transcript. The experimentation was done utilizing the pivotal window, and the incremental comparison algorithm (based on the phrase's size in letters or syllables), were used as algorithms for generating candidate phrases.

In \cite{Campos2018} the sliding window strategy is presented where a set $ S_ {j} $ is generated with a window $v = 1$. The selection of sub phrases is done using a pivot $p_{j}$ and the set $S_{j}$ of candidate sentences is generated by the sub phrases \{${p_{j}, \; p_{j-1} p_{j}, \; p_{j} p_{j + 1}, \; p_{j-1} p_{j} p_{j + 1}}$ \}.

An incremental sub phrase search method was implemented for this article, which is described below:

Let $C \; = \; \{c_{1}, \ldots, c_{n} \}$ the set of $n$ context-specific phrases, and $T = \{t_{1}, \ldots, t_{m} \}$ the original transcript divided into $m$ words, it is intended to build a set $ R = \{(s_1, c_1), \ldots, (s_l, c_l) \}$ formed by pairs $(s_i, c_i)$ such that $c_i$ is an element of the context capable of substituting the segment $s_i = t_{j} ... t_{k} $ for some $ j \leq k$ in $T$.
\begin{algorithm}
	\caption{Candidate incremental search algorithm}
	\label{algo1}
	\begin{algorithmic}[1]
		\Require The context $C = \{ c_{1},\ldots,c_{n} \}$, the transcript $T = \{ t_{1},\ldots,t_{m} \}$, a distance threshold $u$, a distance metric function $d(a, b)$.
		\Ensure a set $R = \{(s_1, c_1), \ldots , (s_l, c_l)\}$ of candidate substitutions.
		\item[] \hspace{-0.6cm}	\hrulefill
		
		\State Calculate maximum phrase size $L_M$ in $C$.
		\State Initialize the set $R = \{\}$
		
		\For {$i$ $\gets$ $1 \ldots m$}
			\State $s$ $\gets$ $t_i$
			\State $j \gets i$
			\While {$j \leq m \And length(s) \leq \frac{L_M}{1-u}$}
				\ForAll {$c \in C \mid length(s)(1-u) \leq length(c) \leq \frac{length(s)}{1-u}$}
					\If {$d(s,c) < u$}
						\State Add the pair $(s, c)$ to the set $R$
					\EndIf
					\State $s \gets s + t_j$
					\State $j \gets j+1$
				\EndFor
			\EndWhile
		\EndFor
		\State \Return $R$
	\end{algorithmic}
\end{algorithm}

The algorithm \ref{algo1} uses a strategy, in which each word $t_i$ in transcript ($T$),  the sub phrase $s$ to be evaluated is incremented word by word until there are no more elements of the context comparable for their size in letters or syllables according to the $u$ threshold. The possible substitutions of $s$ for $c$ are added to the set $R$ as long as their distance is less than $u$. The algorithm's complexity is $O(nm^2)$ where $n$ is the number of elements in the context and $m$ the size of the transcript in words.

\subsection{String Distance Metrics}

The edit distance is used to quantify the difference between two text strings in terms of the number of operations required to transform one string into the other. This work experiments with Levenshtein, Damerau-Levenshtein distance metrics, and Optimal Chain Alignment (OSA).

The Levenshtein distance between two character strings is the number of insertions, deletions, and substitutions required to transform one character string into another \cite{Levenshtein1966}. The Damerau-Levenshtein distance can be intuitively defined as an extension of the Levenshtein distance by adding the transposition of two adjacent characters \cite{Bard2007} as a valid operation. OSA is a restrictive variation of the distance \textit{Damerau-Levenshtein}, where the transpose operation can only be performed once per character \cite{Navarro}, which makes it less computationally expensive.

\subsection{Evolutionary context optimization}

For the generation of contexts, it was decided to use a genetic algorithm constructed from the sentences' transcripts to be corrected. Each individual represents a possible context. For the individuals' construction, all the words were considered individually, and the combinations of 2 words (bigrams) present in the target sentences of the original audios. Individuals were defined by a chromosome where each gene takes the value 1 if the word or bigram is in the context and 0 otherwise.

\begin{algorithm}[H]
	\caption{Context optimization algorithm}
	\label{algo2}
	\begin{algorithmic}[1]
		\Require Population size $N$, number of generations $G$, tournament size $T_{s}$, crossover probability $C_{p}$ and mutation probability $M_{p}$.
        \Ensure The evolved population $p$ and the average generation error $errors$.
		\item[] \hspace{-0.6cm}\hrulefill
		\State $p \gets GenerateInitialPopulation(N)$, $g \gets 0$
		\While {$g < N$}
		\State $errors[g] \gets Evaluate(p)$
		\State $p_{t} \gets Select(p, T_{s})$
		\State $p \gets CrossMutate(p_{t}, C_{p}, M_{p})$
		\State $g \gets g + 1$
		\EndWhile
		\State \Return $p$, $errors$
	\end{algorithmic}
\end{algorithm}

In this way, each individual represents a context, which is a potential parameter to the correction algorithm. To evaluate individuals, the correction algorithm was run with the best combination found in the article \cite{Campos2018} to each of the 451 sentences. The total WER of each context analyzed was returned as a measure of \emph{fitness}. A simple genetic algorithm described in the algorithm \ref{algo2} was used where the function $GenerateInitialPopulation(N)$ produces $N$ individuals randomly. $Evaluate(p)$ is a function that calculates each individual's WER, assigns it as a measure of \emph{fitness} and returns the average WER of the population. The selection was made with a simple tournament strategy. The function $CrossMutate(p_{t}, C_{p}, M_{p})$ performs genetic recombination between the individuals of the population using the random crossing point technique shown in the algorithm \ref{algo3}. Subsequently, the individual-to-individual and gene-to-gene mutation process was carried out according to the mutation probability value, which was reduced every ten generations to reduce the fluctuation and stabilize the error when it approached a minimum.

\begin{algorithm}[H]
	\caption{Crossing operation between individuals}
	\label{algo3}
	\begin{algorithmic}[1]
		\Require Individuals $I_{1}$ and $I_{2}$ to be combined, a crossing point $c_{i}$, and a chromosome size $c_{size}$
        \Ensure Ancestry of input individuals defined as $H_{1}$ and $H_{2}$.
        \State $H_{1} \gets g1_{1}g1_{2}...g1_{c_{i}}g2_{c_{i} + 1}g2_{c_{i} + 2}...g2_{c_{size}}$
        \State $H_{2} \gets g2_{1}g2_{2}...g2_{c_{i}}g1_{c_{i} + 1}g1_{c_{i} + 2}...g1_{c_{size}}$
		\State \Return $H_{1}$, $H_{2}$
	\end{algorithmic}
\end{algorithm}

\section{Experiments}
The present work experiments were carried out using 451 phrases transcribed by the \emph{Google} speech recognition system. This same corpus was used in \cite{Campos2018}, where details of the said corpus' collection process and format can also be obtained.

\subsection{Corrector setting variants} \label{sub:config_variants}

To compare the variants of each element of the algorithm, a total of 72 experiments were carried out with all the combinations of the methods presented in table ~\ref{tab:grid}.

\begin{table}
	\center
	\caption{Method variants for each element of the algorithm.}
	\label{tab:grid}	
	\begin{tabular}{| c | c | c | c |}
		\hline
		\textbf{Representation} & \textbf{Phrases Generation} & \textbf{Distance metric} & \textbf{Google STT} \\
		\hline
	    Simple text & WIN & Levenshtein & Basic \\
		IPA & LET & OSA & Contextual \\
		DM & SYL & Damerau-Levenshtein & \\
		DMV & & & \\
		\hline		
	\end{tabular}
\end{table}

The text of the phrase recognized by the STT system and the context's phrases were represented differently for processing. For all text representations, it was necessary to carry out a normalization process to remove some punctuation symbols and characters. This normalized version was used directly in the form of simple text or transformed into one of the analyzed phonetic representations: International Phonetic Alphabet (IPA), Double Metaphone (DM), or Double Metaphone with vowels (DMV).

The generation of candidate sentences was experimented with using the pivot window methods (WIN) and the incremental comparison according to the number of characters (LET) or syllables (SYL).

Levenshtein distance and its variants OSA and Damerau-Levenshtein were used as editing distance metrics. Each combination was tested using as input data the 451 transcripts obtained by using the basic \emph{Google} method and subsequently the transcript resulting from sending the context reported in \cite{Campos2018} to the \emph{Google} service.

For each of the 72 different configurations, different confidence thresholds of the editing metrics were experimented with increments of 0.05 up to a maximum of 0.6. The evaluation of each experimental setup was carried out using the globally accumulated WER metric calculated from the number of edits required to transform the hypothetical transcript into the correct sentence for each example.

\subsection{Context optimization} \label{sec:context_optimization}

The experimentation described in the previous section was carried out with an empirically generated context according to the a priori knowledge of the domain phrases where transcription errors were observed. In order to optimize the context, a genetic algorithm was run whose parameters were calibrated by conducting experimentation of 30 executions with a reduced version of the problem using a chromosome of size 50, yielding the best results with a population of 50 individuals, 100 generations, 95\% probability of crossover and 5\% probability of mutation.

Once the parameters had been calibrated, the context was optimized using a chromosome size of 355. Each gene represents one of the words or bigrams present in the transcripts of the audios used in the experimentation. The mutation factor was reduced by 20\% every 10 generations. Individuals were evolving for 100 generations.

As a function of \emph{fitness}, the total WER obtained when executing the correction algorithm for the simple transcription of \emph{Google} was used using the individual defined by the chromosome as context. The phonetic corrector was run using IPA representation, pivot window selection, Levenshtein distance, and threshold of 0.4, which was the best configuration reported in \cite{Campos2018}.

Five evolutionary processes of 100 generations were executed where the population was initialized with the 25 best individuals from the previous round and 25 randomly generated to explore different evolutionary variants. The experimentation was done with an Intel i7 processor, 8GB of RAM, and a Debian GNU/Linux operating system. The algorithm was implemented in Python3 and ran five times for a total of 70 hours.

\subsection{Error correction with optimized context} \label{sub:finalexp}

In this phase of the experimentation, tests were carried out to measure the effects of context on speech recognition and subsequent correction. The previously generated audio files were sent back to the recognition system in its basic mode and with the genetically generated context. This gave us a direct comparison of the effect of using the optimized context concerning the transcript obtained without sending context; besides, it gave us two baselines on which to apply the correction process to the new transcripts.

To compare the correction process results, the algorithm was applied to the new transcripts produced by both versions of the \emph{Google} recognizer. The four configuration variants that presented the best results were used in the experimentation described in section \ref{sub:config_variants}. Taking as input the transcripts obtained by the two modalities of the \emph{Google} service, two experiments were executed for each one of them. In the first, the experimental context used in \cite{Campos2018} was used, which we will denote as $C_m$, and in the second experiment, the new genetically generated context $C_g$.

\section{Results}

\begin{figure}[H]
	\subfloat[Text representation] {\includegraphics[width=0.5\textwidth]{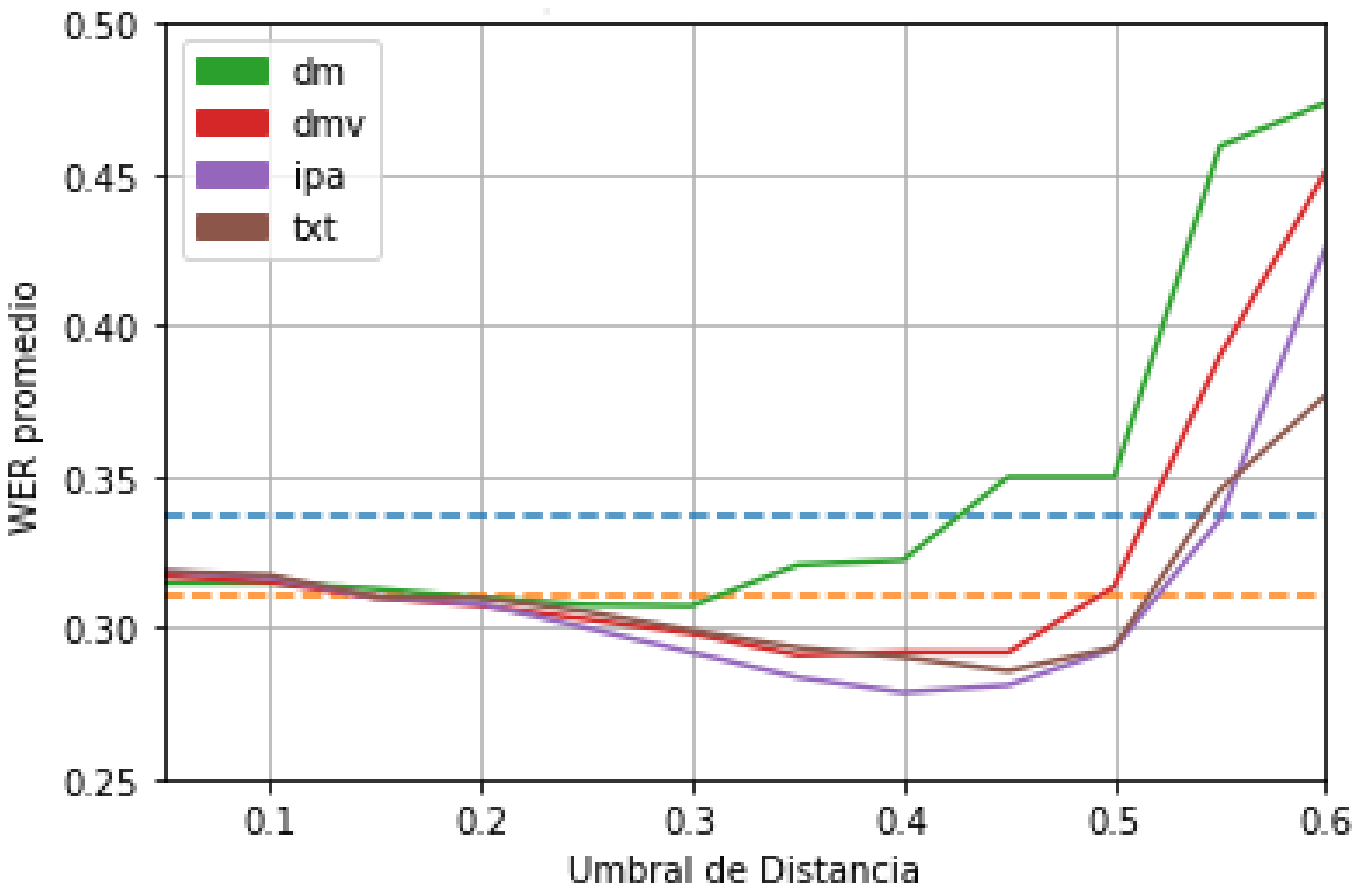}}	\label{fig:rep}
	\subfloat[Candidates generation] {\includegraphics[width=0.5\textwidth]{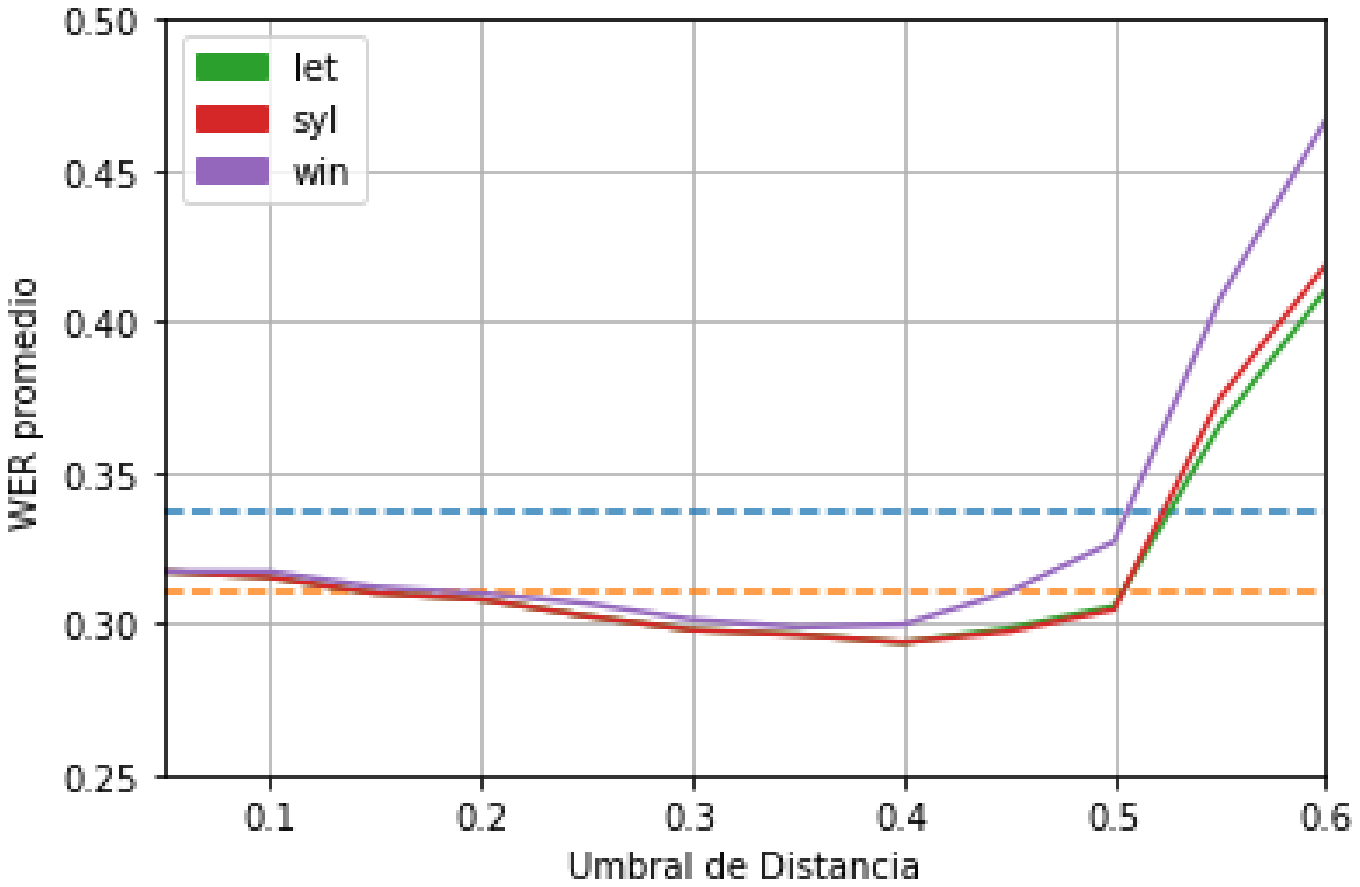}} \label{fig:gen}
	\caption{Average WER for different text representations (a) and candidate generation methods (b)}
	\label{fig:mode_search}
\end{figure}

 Fig. ~\ref{fig:mode_search} (a) shows the variation in the average WER obtained in the experiments grouped by the representation mode with different distance thresholds. The horizontal lines correspond to the WER obtained with the transcription of the basic STT (33.7\%) and the contextual STT (31.1\%). The effect of reducing the WER is observed when transforming the plain text into IPA, especially around the value 0.4 for the threshold, where a minimum average WER of 27.8\% is reached. On the other hand, the representation in DM produced similar results to the other methods for small values of the threshold, however around 0.3, its corrective capacity decreases consistently. The DMV version sustains its performance on par with plain text up to a threshold of 0.4.

In Fig. ~\ref {fig:mode_search} (b), the average WER obtained by the different experimental configurations grouped by the candidate phrase generation algorithm is observed. The graph shows a better performance of the incremental comparison variants, either by size in letters or syllables, compared to the pivot window. The minimum average WER (29.4\%) is reached with a threshold of 0.4 for the LET method. The SYL version shows very similar results; however, the computational cost of processing is higher.

The best results were obtained with the IPA representation configuration and the LET candidate selection method. The WER obtained from the basic transcript decreased from 33.7\% to 28.1\%, and for the contextual transcript, it decreased from 31.1\% to 27.3\%. This configuration presented a global reduction in the relative WER of 19.3\%. Concerning the three distance metrics evaluated, the difference in the results was practically nil.

The results obtained from the experimentation with the genetic algorithms for the optimization of the context using the experimental configuration described in the section \ref{sec:context_optimization}, indicate that an average error was obtained in the fifth execution of the experiment of 26.5\% , which started with an average error of 31.2\% and decreased to an average of 24.9\% in generation 100.

\begin{figure}[H]
	\centering
	\includegraphics[width=0.5\textwidth]{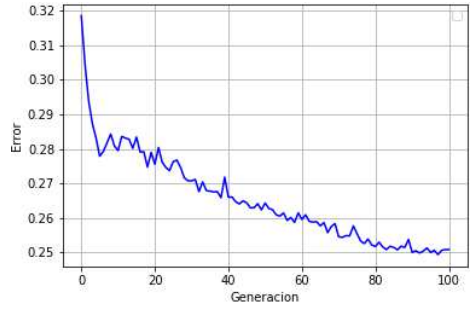}
	\caption{Graph of the average error per generation of the genetic algorithm}
	\label{fig:error_ag}
\end{figure}

In Fig. ~\ref{fig:error_ag} the average errors of the 100 generations in the fifth run of the experiment are shown. The best context found with this strategy contained 64 unigrams and 117 bigrams with a total WER of 24.7\%.

\begin{figure}[H]
	\subfloat [Pivot window - Basic STT] {\includegraphics[width = 0.5 \textwidth]{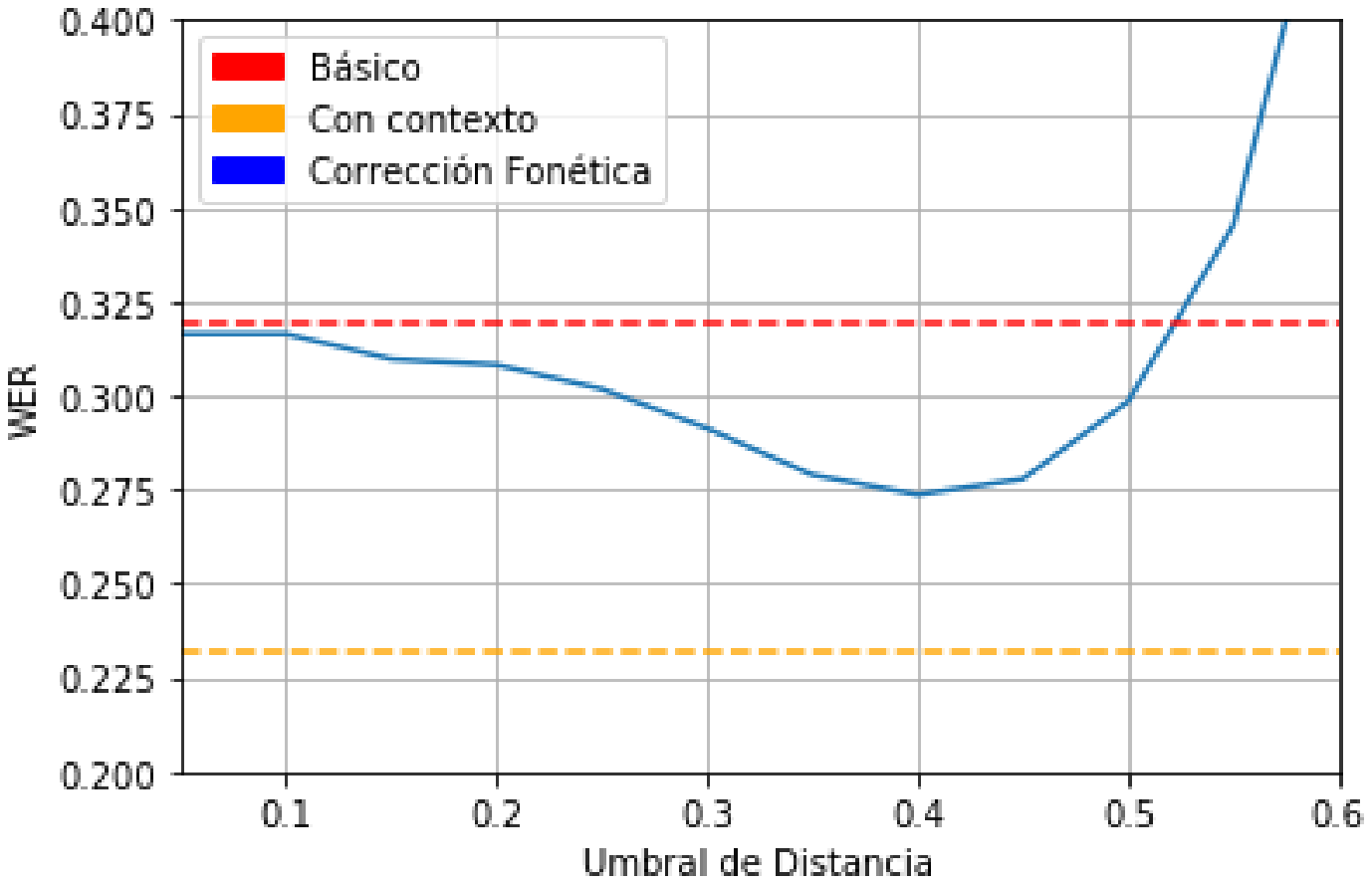}}
    \subfloat [Character size - Basic STT] {\includegraphics[width = 0.5 \textwidth]{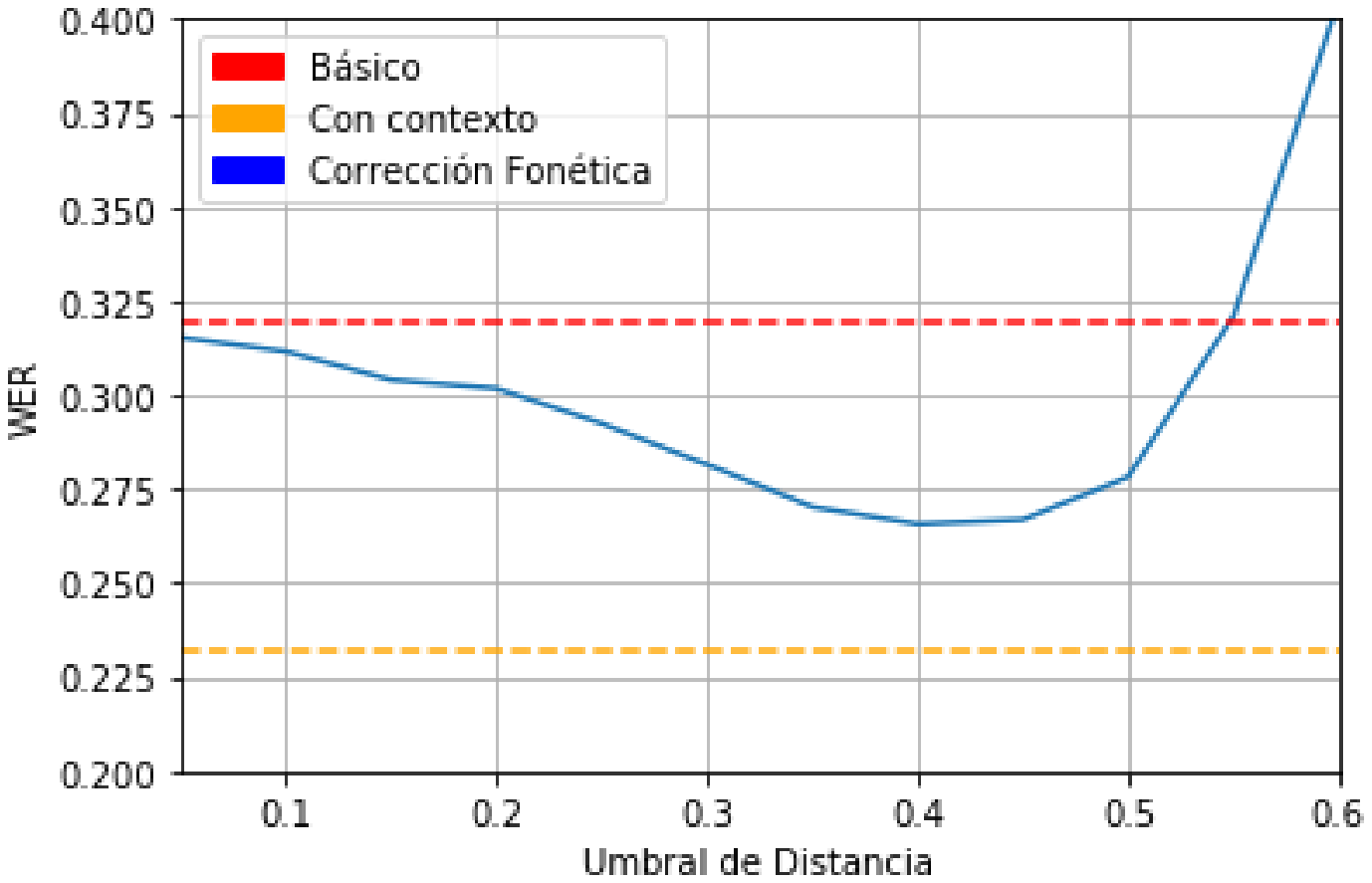}} \\
    \subfloat [Pivot Window - Contextual STT] {\includegraphics[width = 0.5 \textwidth]{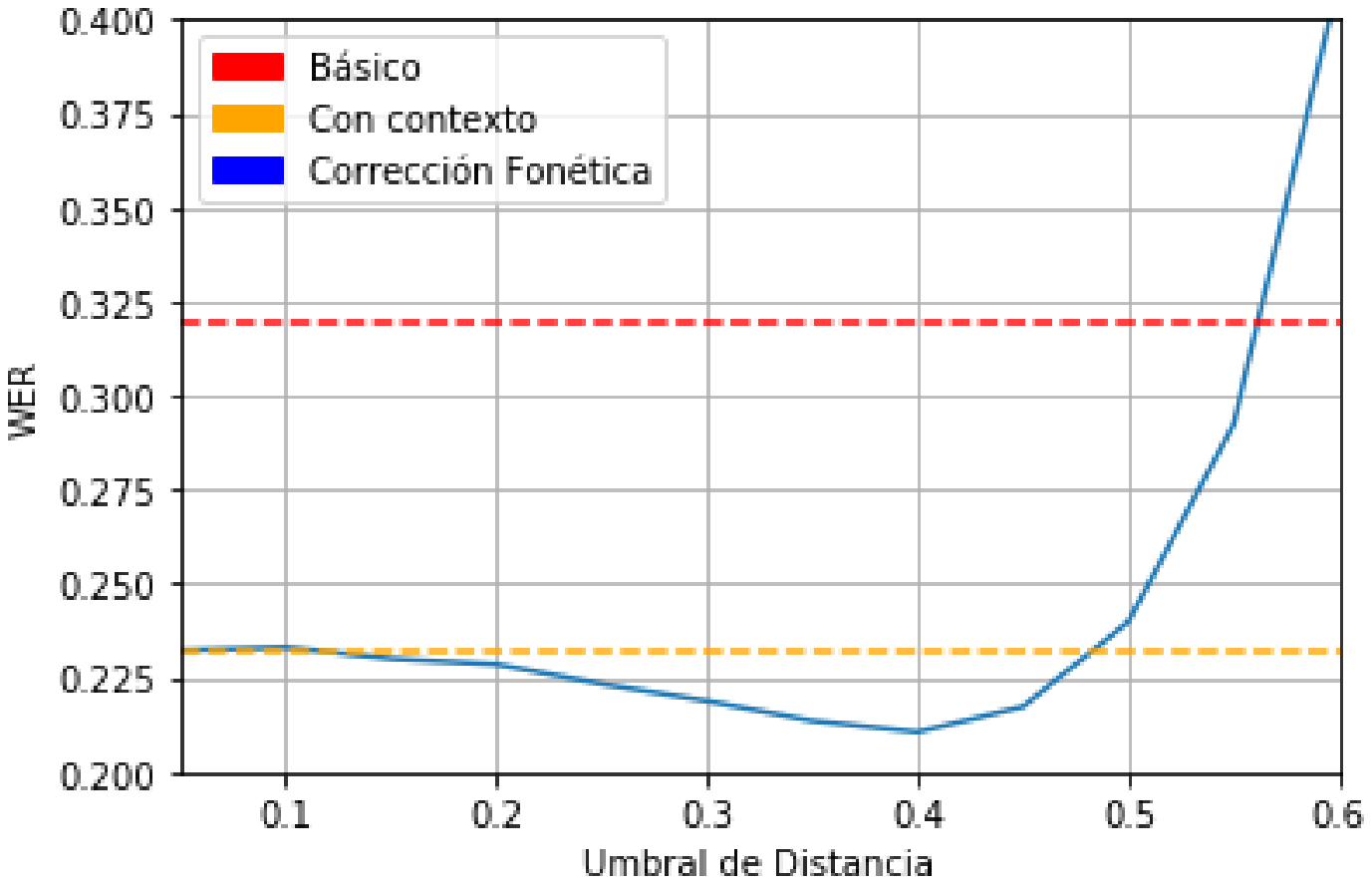}}
    \subfloat [Character size - contextual STT] {\includegraphics [width = 0.5 \textwidth]{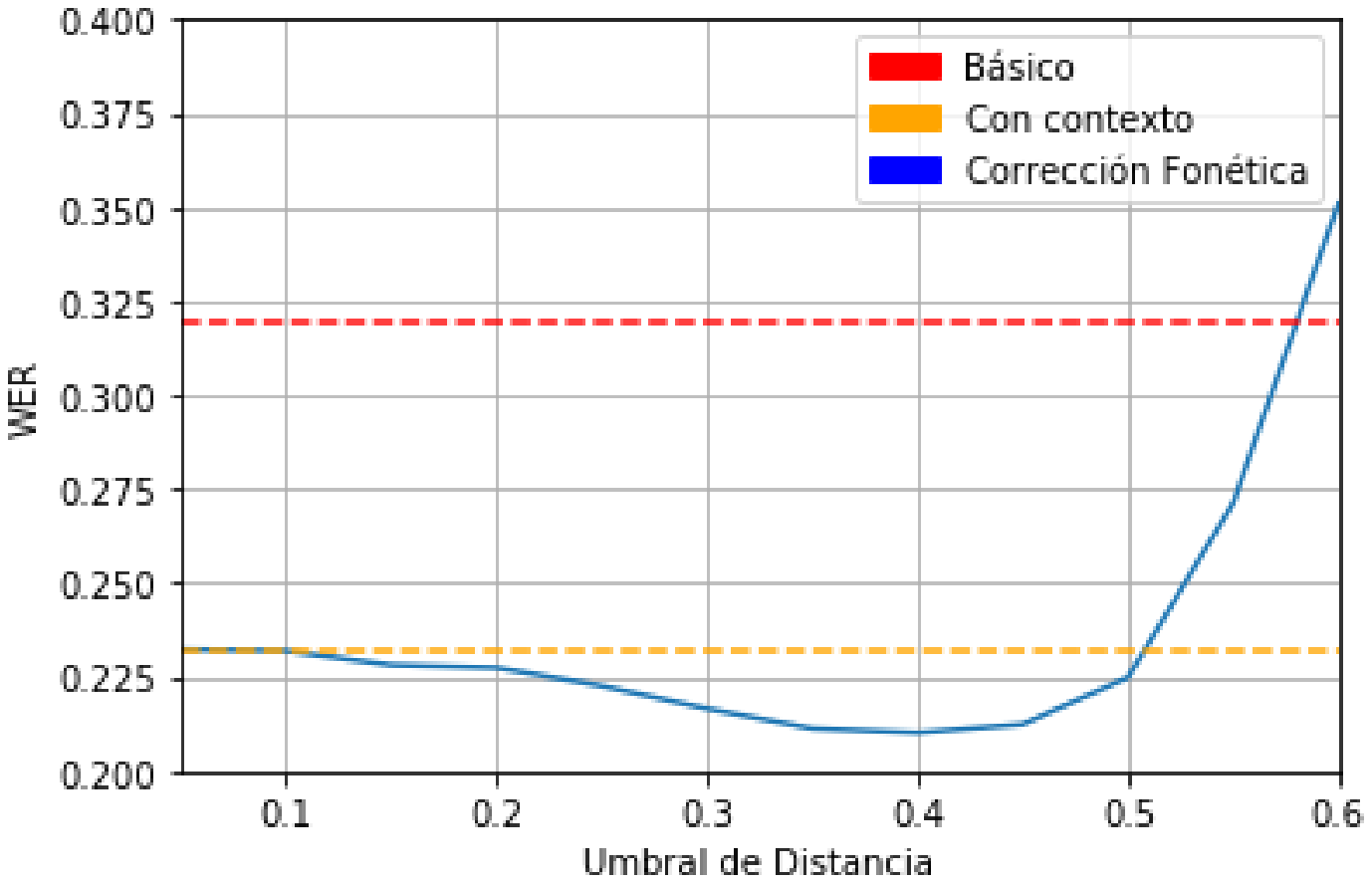}}
	
	\caption{Results using $C_m$ as input to the correction algorithm}
	\label{fig:g19_c18}
\end{figure}

In the final phase of the experimentation described in section \ref{sub:finalexp}, two baselines are obtained to execute the correction process. The WER obtained when comparing the actual sentences spoken with the basic transcription was 32.0\%, while incorporating the genetically generated context, the WER was considerably reduced to 23.2\%. This result allows us to see the impact that an optimized context has on the language model used by \emph{Google} by reducing the relative WER by 27.3\%.

Fig. ~\ref {fig:g19_c18} shows the correction algorithm results using the IPA representation with the context $C_m$. Starting from the basic STT, the minimum WER is obtained with a threshold of 0.4 and the LET selection process. With this configuration, the total WER is reduced from 32.0\% to 26.6\%, representing a reduction of 16.9\% in the WER relative. When starting the correction process from the transcription of the contextual STT, the WER decreased from 23.2\% to 21.0\%, reaching a 9.5\% improvement in the relative WER.

\begin{figure}[H]
	\subfloat [Pivot Window - Basic STT] {\includegraphics [width = 0.5 \textwidth]{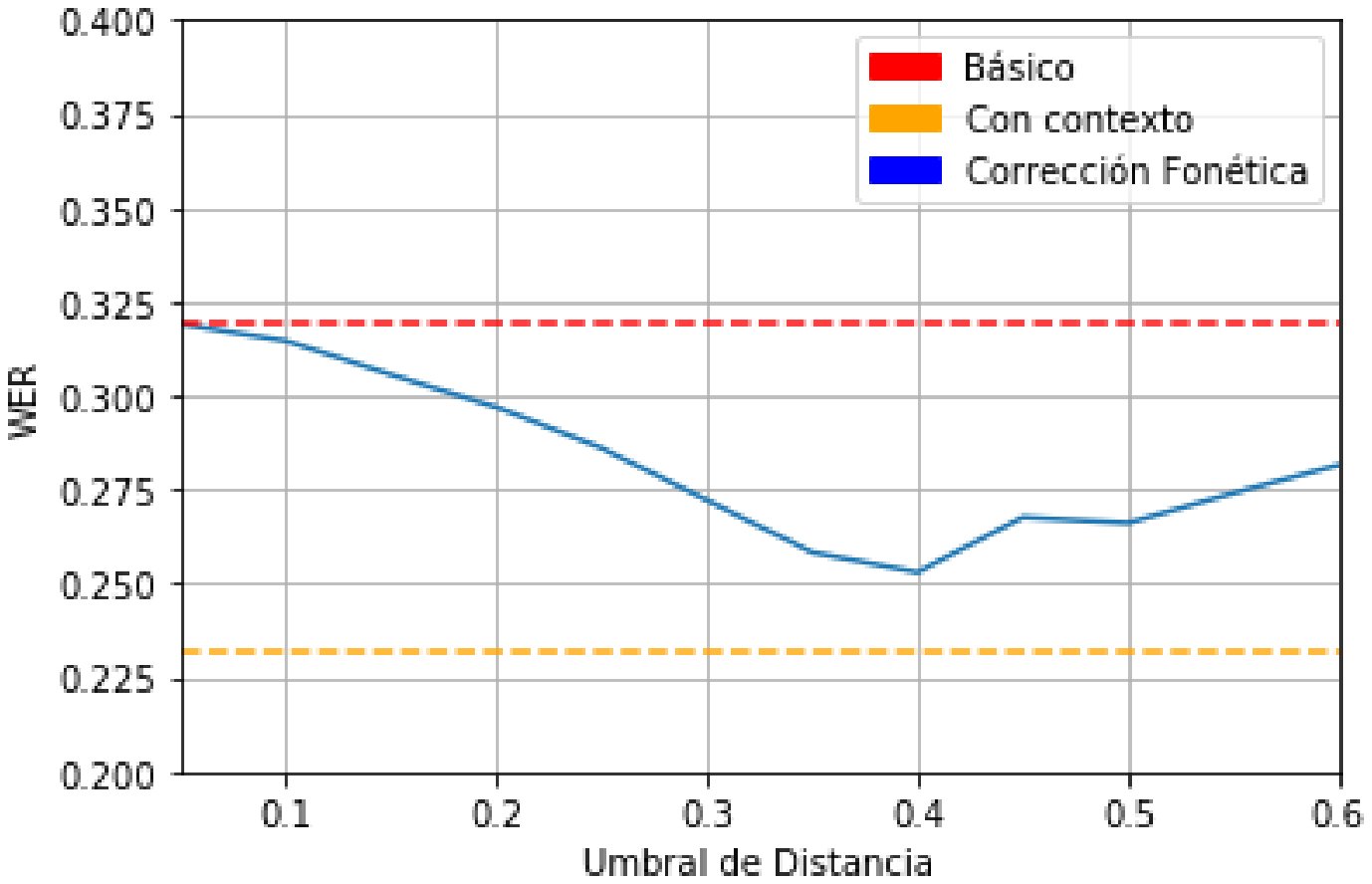}}
    \subfloat [Character size - Basic STT] {\includegraphics [width = 0.5 \textwidth]{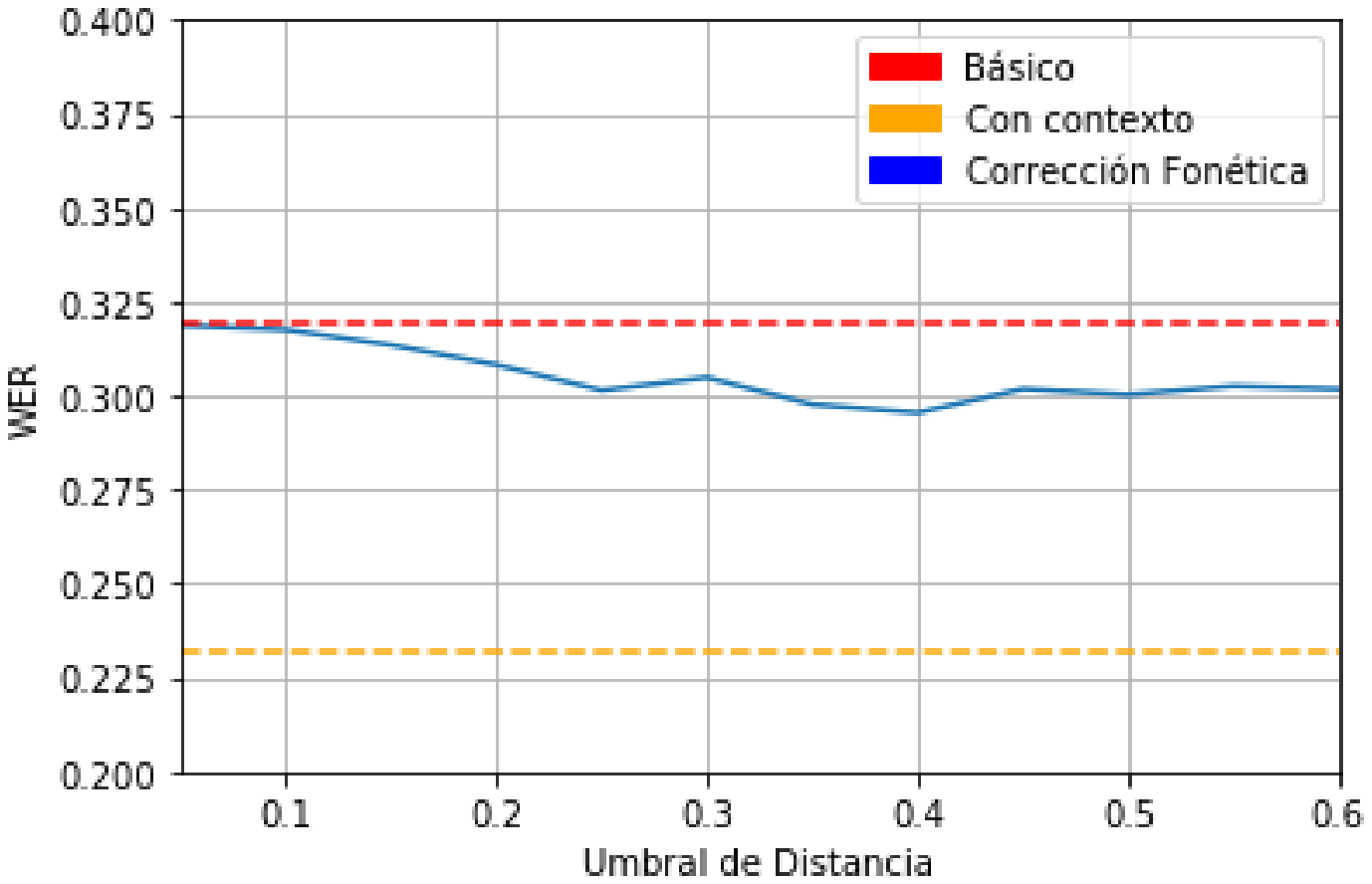}} \\
    \subfloat [Pivot window - Contextual STT] {\includegraphics [width = 0.5 \textwidth]{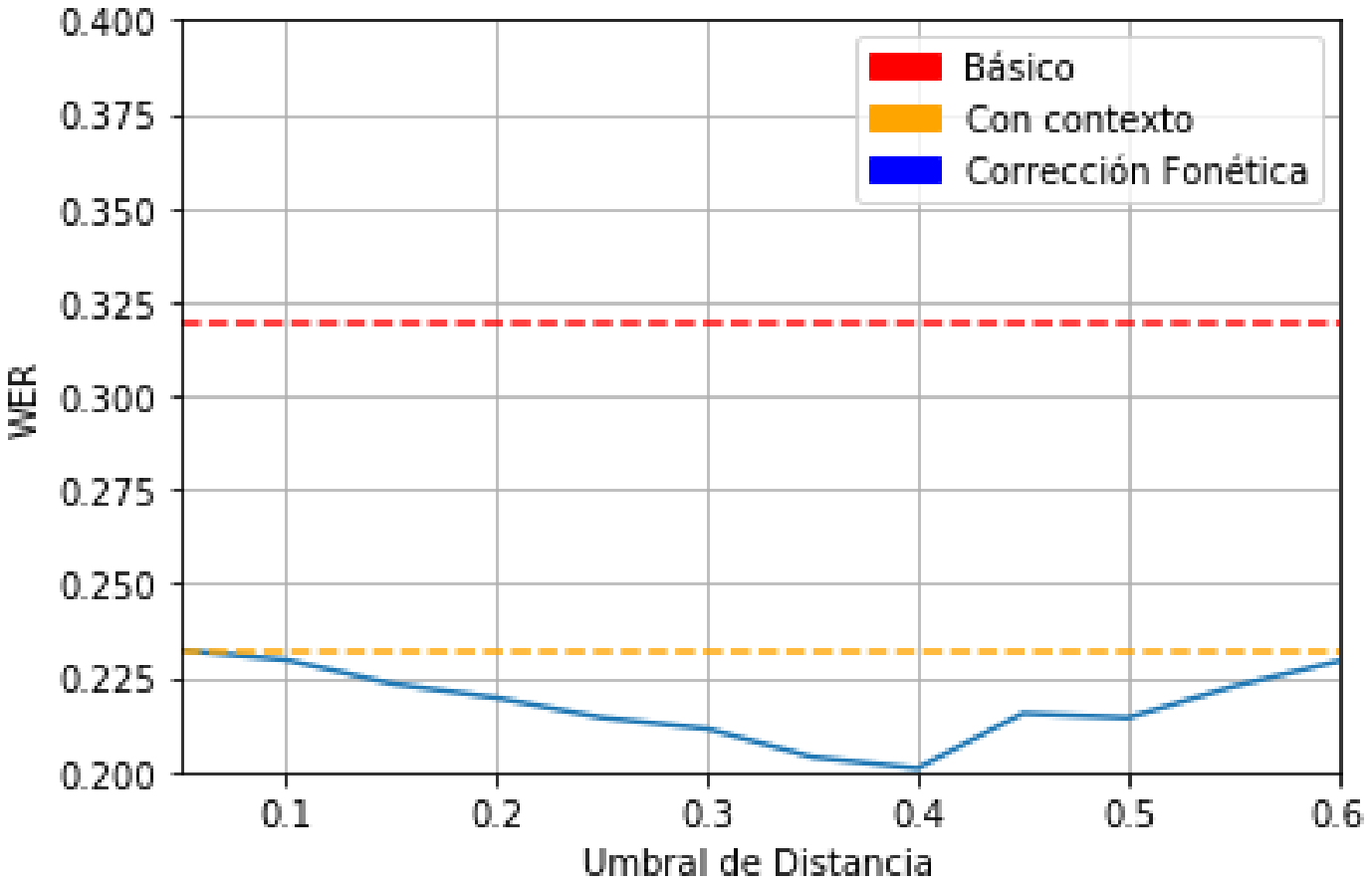}}
    \subfloat [Character size - contextual STT] {\includegraphics [width = 0.5 \textwidth]{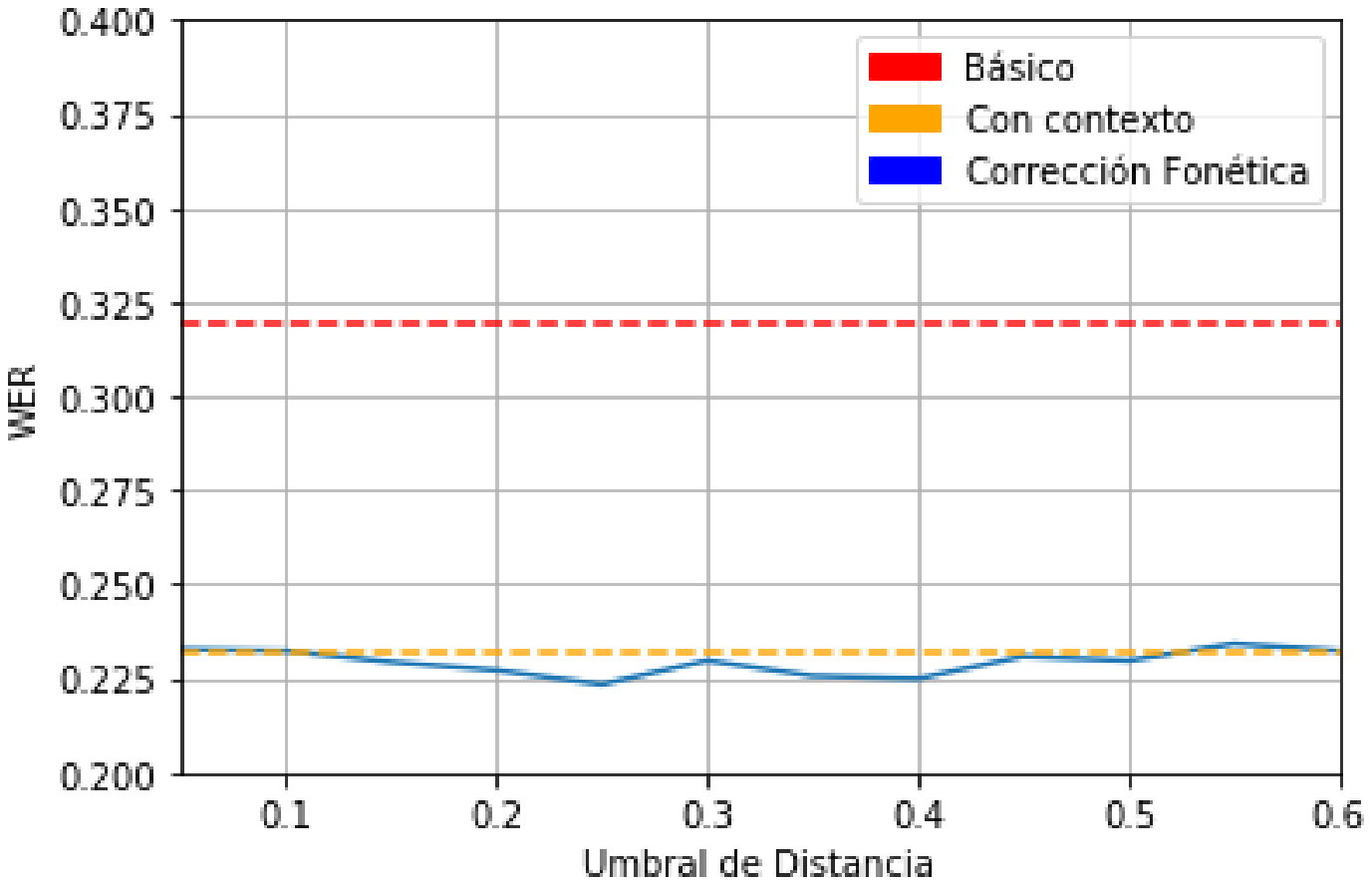}}
    
    \caption{Results using $ C_g $ as input to the correction algorithm}
	\label{fig:g19_c19}
\end{figure}

Similarly, Fig. ~\ref{fig:g19_c19} shows results using the genetically generated context $C_g$ as input to the correction algorithm. Using this context, a reduction of the minimum WER when using the WIN candidate generation procedure is observed. However, it does not seem to generate good results with the LET method. The absolute minimum WER for the basic STT is 25.3\%, thus reducing the relative WER of 21.0\%. Starting from the contextual STT, a minimum of 20.1\% is reached, representing a reduction of 13.6\% in relative WER.

\section{Conclusions and future work}

From the results obtained in the experimentation, the phonetic correction algorithm's usefulness is shown to reduce errors in the transcription of \emph{Google}, both in its basic and contextual versions. It is observed that the best configuration for the algorithm is obtained using IPA as phonetic representation and incremental selection by letters, managing to reduce the relative WER by 19.0\%.

Similarly, we can mention that genetic algorithms are an efficient alternative for generating contexts since they managed to reduce the WER of \emph{Google} basic transcription from 32.0\% to 23.2\%. The context was shown to have a crucial value in the performance of the algorithm.

The best results were obtained from the combination of the phonetic correction with the evolutionary optimization of the context, achieving a reduction of the absolute WER of 11.9\% by decreasing it from 32.0\% to 20.1\%, representing an improvement in relative WER of 37.2\%.

The fact that both the phonetic correction algorithm and the evolutionary context optimization are independent of the system used for the transcription and application domain means that the strategy presented can be extended to different ASR systems and application domains.

The algorithms presented throughout this article can take advantage of a priori knowledge of the application domain to mitigate the cold start problem. The above is because if the initial transcripts are not available, a context generated with human knowledge of the domain can be used, as in \cite{Campos2018}, which can be complemented with genetic algorithms as information is collected about interactions with actual users of the system.

Among future research lines, it is necessary to validate the results with a corpus of different application domains; in addition, experimentation using weighted editing costs considers phonetic characteristics of Spanish and the original audio such as noise, duration, the energy of the signal, among others. Another line of research is the comparison with \emph{deep learning} algorithms since the problem of error correction in ASR systems can be considered a translation of erroneous transcripts to correct transcripts, so \emph{algorithms Machine translation} can be helpful.

\bibliographystyle{splncs04}
\bibliography{reference}

\end{document}